# Propagating Wave Patterns in a Derivative Nonlinear Schrödinger System with Quintic Nonlinearity

**Colin ROGERS**[(1)], **Boris MALOMED**[(2)], **Jin Hua LI**[(3)], and

**Kwok Wing CHOW**\*[(3)]

(1) = Australian Research Council Centre of Excellence for Mathematics & Statistics of Complex Systems, School of Mathematics,

The University of New South Wales, Sydney, NSW 2052, Australia

(2) = Department of Physical Electronics, School of Electrical Engineering, Faculty of Engineering, Tel Aviv University, Tel Aviv 69978, Israel

(3) = Department of Mechanical Engineering, University of Hong Kong, Pokfulam, Hong Kong

\* = Corresponding author

**Email:** kwchow@hku.hk      **Fax:** (852) 2858 5415

**Abstract**

Exact expressions are obtained for a diversity of propagating patterns for a derivative nonlinear Schrödinger equation with a quintic nonlinearity. These patterns include bright pulses, fronts and dark solitons. The evolution of the wave envelope is determined via a pair of integrals of motion, and reduction is achieved to Jacobi elliptic cn and dn function representations. Numerical simulations are performed to establish the existence of parameter ranges for stability. The derivative quintic nonlinear Schrödinger model equations investigated here are important in the analysis of strong optical signals propagating in spatial or temporal waveguides.

# 1. Introduction

The nonlinear Schrödinger (NLS) equation is a widely studied model applicable to many disciplines in engineering and applied science.[1–7] Nonlinear optics is one of the fields where the NLS equations are profoundly important models. Self trapping of light and formation of spatial, temporal and spatiotemporal solitons have been studied intensively in photonics. In this context, the evolution of slowly varying electromagnetic field is governed by a NLS equation. Stable solitons are supported by the competing effects of diffraction and self-focusing (Kerr) nonlinearity in the context of spatial waveguides, or group velocity dispersion and the same nonlinearity in the analysis of the propagation of temporal pulses along fibers.[6,7]

The one dimensional NLS equation which includes only dispersion and cubic nonlinearity is exactly integrable.[1] Higher order nonlinear effects need to be restored in many circumstances. In fluid mechanics, such considerations arise in nonlinear water waves[8,9] and shear flow instability.[2] In optics, experimental evidence reveals that quintic effects are relevant if a sufficiently strong optical field is allowed to propagate in a material medium. Cubic-quintic nonlinearities can be observed in composite optical media (colloids), organic materials, and special glasses.[10–13] Theoretically, the refractive index can be modeled in terms of even powers of the amplitude of the electric field,[14] and exact solutions can be found.[15] Nonlinear Schrödinger models with cubic-quintic, nonlinearity have been considered with various additional physical effects, e.g. fourth order dispersion,[16] non-Kerr media,[17] and presence of energy gain/loss, i.e. Ginzburg Landau models.[18] Furthermore, super Gaussian solitons in semiconductor doped glass fibers, a dispersive medium with cubic-quintic nonlinearity, can be dynamically stable.[19] It is remarked that cubic-quintic NLS models also arise in elasticity.[20]

The stability of nonlinear waves can be dramatically altered by fifth order nonlinear terms. Weak quintic nonlinearity can stabilize one type of periodic waves, but destabilizes others,[21] depending on whether the medium is focusing or defocusing.

Exact solutions for NLS equations with cubic-quintic nonlinearity have attracted considerable attention.[15,22–28] In this context, the modulus of the electric field can typically be expressed in terms of the classical Jacobi elliptic functions.[22,23,29] Moreover, several families of exact analytical solutions can be obtained using elementary functions.[24–28] The results may be extended to inhomogeneous media.[30]

Derivative nonlinear Schrödinger (DNLS) equations constitute another class of physically important evolution systems. Thus, additional physics is brought into play when the group velocity changes with the light intensity. In practice, such 'self-steepening' effects[7] create asymmetry and distort the wave forms. The effect of self-steepening on localized nonlinear modes has attracted attention recently, and, in particular, optical solitons coupled to a continuous background have been studied by a generalized Lax pair technique.[31] Furthermore, self steepening permits the occurrence of 'anti-dark' solitons. Both dark and anti-dark solitons can co-exist on the same background.[32] Effects of polarization, higher order dispersion and other optical phenomena on ultrashort pulses have been investigated.[33] In addition, self-steepening can play a role in the modulation instabilities of continuous waves[34] and supercontinuum generation.[35] Several versions of DNLS equations have been studied in the literature. The self-steepening term that arises in the optical setting is commonly associated with the Kaup-Newell equation.[36] Here, it proves more convenient to employ the Chen-Lee-Liu variant of the DNLS equation.[37] These two versions are related by a gauge transformation.

The goal of the present work is to study the propagation of an electric field in a NLS model with both quintic nonlinearity and self-steepening,[7] with the latter represented by the Chen-Lee-Liu equation.[37] A novel procedure is presented to obtain exact analytic representations for propagating patterns described by cubic-quintic derivative nonlinear Schrödinger equation. These exact solutions represent, in particular, bright pulses, dark solitary waves, fronts/shocks and periodic patterns. The method involves the exploitation of two integrals of motion (H, ℑ) obtained via the coupled nonlinear system resulting from the introduction of a wave-packet ansatz into the class of derivative NLS equations. In that sense, the procedure complements that involving the exact solution of Hamiltonian Ermakov-Ray-Reid systems via their pairs of integrals of motion.[38,39] The stability of the exact solutions for localized modes is explored by means of direct simulations of the perturbed evolution of the solutions.

## 2. Formulation

Here, we shall be concerned with the cubic-quintic nonlinear Schrödinger equation in the form

$$i\,\Psi_t + \hat{\lambda}\,\Psi_{xx} + \hat{\mu}\,|\Psi|^2\,\Psi + i\,\hat{\alpha}\,|\Psi|^2\,\Psi_x + \hat{\nu}\,|\Psi|^4\,\Psi = 0. \tag{1}$$

The slowly varying envelope $\Psi$ evolves under the influence of dispersion (as measured by $\hat{\lambda}$), cubic ($\hat{\mu}$) and quintic ($\hat{\nu}$) nonlinearities, while the $|\Psi|^2\Psi_x$ term is commonly associated with 'self steepening'. In hydrodynamics, the coordinates $t$ and $x$ are typically slow time and spatial coordinate moving with the group velocity.

In an optical fiber setting, these denote distance and retarded time, while for spatial waveguides, they denote the propagation coordinate and diffraction respectively.

It is relevant to mention that eq. (1) is invariant with respect to a Galilean transformation, which, in contrast to the usual NLS equation, involves a change of coefficients of the equation. More precisely, if $\Psi(x, t)$ is a solution of eq. (1), then its boosted counterpart, moving at arbitrary velocity $c$,

$$\tilde{\Psi} = \Psi(x-ct,t)\exp\left(\frac{ic}{2\hat{\lambda}}x - \frac{ic^2}{4\hat{\lambda}}t\right),$$

is a solution to eq. (1) with the shifted cubic coefficient:

$$\hat{\mu} \to \hat{\mu} - c\hat{\alpha}/\left(2\hat{\lambda}\right) \ .$$

Returning to the search for exact solutions, propagating patterns of eq. (1) are now sought via the wave packet ansatz

$$\Psi = [\phi(x-\mu t) + i\psi(x-\mu t)]\exp[i(v x - \lambda t)], \tag{2}$$

where $\mu$ denotes the (constant) travelling speed of the envelope.

Introduction of representation eq. (2) into eq. (1) produces two coupled nonlinear ordinary differential equations for $\phi$ and $\psi$. This pair admits two integrals of motion, namely

$$\dot{\phi}\psi - \dot{\psi}\phi = \Im + \left(v - \frac{\mu}{2\hat{\lambda}}\right)\Sigma + \frac{\hat{\alpha}}{4\hat{\lambda}}\Sigma^2 , \tag{3}$$

and

$$\dot{\phi}^2 + \dot{\psi}^2 = 2\mathsf{H} \ +\left(v^2 - \frac{\lambda}{\hat{\lambda}}\right)\Sigma + \left(\frac{\hat{\alpha}\,v - \hat{\mu}}{2\,\hat{\lambda}}\right)\Sigma^2 - \frac{\hat{v}}{3\,\hat{\lambda}}\Sigma^3 , \qquad (4)$$

where

$$\Sigma = \phi^2 + \psi^2 \qquad (5)$$

and $\Im$ and $\mathsf{H}$ are integration constants. Here eq. (4) corresponds to a Hamiltonian invariant. In the above, the dot denotes the derivative with respect to the propagating wave phase variable $x - \mu t$.

On use of the identity

$$\left(\phi^2 + \psi^2\right)\left(\dot{\phi}^2 + \dot{\psi}^2\right) - \left(\phi\,\dot{\phi} + \psi\,\dot{\psi}\right)^2 = \left(\phi\,\dot{\psi} - \psi\,\dot{\phi}\right)^2 , \qquad (6)$$

it is readily shown that the evolution of $\Sigma$ is given by:

$$\dot{\Sigma}^2 = \Sigma\left\{8\mathsf{H} \ + 4\left(v^2 - \frac{\lambda}{\hat{\lambda}}\right)\Sigma + 2\left(\frac{\hat{\alpha}\,v - \hat{\mu}}{\hat{\lambda}}\right)\Sigma^2 - \left(\frac{4\,\hat{v}}{3\,\hat{\lambda}}\right)\Sigma^3\right\} \\ - 4\left[\Im + \left(v - \frac{\mu}{2\,\hat{\lambda}}\right)\Sigma + \frac{\hat{\alpha}}{4\,\hat{\lambda}}\Sigma^2\right]^2 . \qquad (7)$$

To recover $\Psi$, it is convenient to introduce the auxiliary variables $\Delta$ and $\Theta$ given by

$$\Delta = \frac{\phi}{\psi}, \quad \Theta = \tan^{-1}\Delta = \tan^{-1}\left(\frac{\phi}{\psi}\right) , \qquad (8)$$

whence

$$\Theta = \int^{x-\mu t} \frac{\Im + \left(v - \dfrac{\mu}{2\hat{\lambda}}\right)\Sigma + \left(\dfrac{\hat{\alpha}}{4\hat{\lambda}}\right)\Sigma^2}{\Sigma} d\xi ,  \qquad (9)$$

where $\xi$ is a dummy variable of integration. The corresponding class of exact solutions of the NLS equations (eq. (1)) is then given by

$$\Psi = \Sigma^{1/2} \exp[-i\Theta + i(vx - \lambda t)] ,  \qquad (10)$$

where $\Sigma$ (eq. (5)), $\Theta$ are determined from eqs. (7, 9) respectively.

## 3. Examples of Propagation Patterns

The procedure outlined in Section 2 shows that in general the modulus $\Sigma^{1/2}$ can, by virtue of eq. (7), be expressed in terms of elliptic functions. On appropriate choice of the elliptic functions or hyperbolic reductions, various patterns of physical interests can be identified. In particular, exact solutions known earlier in the literature are special cases of eq. (7). Here, we restrict attention to patterns involving the Jacobi elliptic function dn, as it has no zeros. Periodic patterns can, in principle, also be obtained in terms of the Jacobi elliptic functions cn and sn, but they may possess singularities as sn and cn pass through zero. Here, the following important integrals involving dn will be called upon:

$$\int \mathrm{dn}(z) \mathrm{d}z = \sin^{-1}[\mathrm{sn}(z)] , \qquad \int \frac{\mathrm{d}z}{\mathrm{dn}(z)} = -\left(\frac{1}{\sqrt{1-k^2}}\right) \sin^{-1}\left[\frac{\mathrm{cn}(z)}{\mathrm{dn}(z)}\right] .$$

**Periodic Solutions**

(i) The function

$$\Sigma = A_0 \text{dn}(x - \mu t)$$

satisfies

$$\dot{\Sigma}^2 = A_0^2 (k^2 - 1) + (2 - k^2)\Sigma^2 - \frac{\Sigma^4}{A_0^2},$$

where $k$ is the modulus of the elliptic function, $A_0$ denotes the amplitude, and 'overhead dot' indicates a derivative with respect to the argument $x - \mu t$. Alignment with eq. (7) shows that eq. (1) admits a particular exact solution

$$\Psi = \sqrt{A_0 \text{dn}(x - \mu t)} \exp\left\{ \frac{i \mu x}{2\hat{\lambda}} - i\left[\frac{\mu^2}{4\hat{\lambda}} - \frac{\Im \hat{\alpha}}{2} - \hat{\lambda}\left(\frac{2 - k^2}{4}\right)\right]t \right.$$

$$\left. - \frac{i \hat{\alpha} A_0 \sin^{-1}[\text{sn}(x - \mu t)]}{4\hat{\lambda}} + \left(\frac{i \Im}{A_0 \sqrt{1 - k^2}}\right) \sin^{-1}\left[\frac{\text{cn}(x - \mu t)}{\text{dn}(x - \mu t)}\right] \right\},$$

(11)

where the amplitude $A_0$, the phase speed $\mu$, and the parameter $\Im$ are given, in turn, by

$$\frac{4 \hat{\nu}}{3 \hat{\lambda}} + \frac{\hat{\alpha}^2}{4 \hat{\lambda}^2} = \frac{1}{A_0^2},$$

$$\mu = \frac{2\hat{\lambda}\hat{\mu}}{\hat{\alpha}}, \quad \mathfrak{I} = \left(\frac{\sqrt{1-k^2}}{2}\right)A_0. \tag{12}$$

Necessary conditions for the existence of these patterns can be obtained from the latter relations. Thus, this wave pattern can only exist if self-steepening is present ($\hat{\alpha} \neq 0$), while the quintic nonlinearity can in principle be either positive or negative.

The fact that exact solution eq. (11) is available at the *single value* of the velocity, given by the second relation in eq. (12) (the same property is admitted by other exact solutions, see below), is explained by the specific form of the Galilean invariance of eq. (1), which involves the shift of the cubic coefficient $\hat{\mu} \to \hat{\mu} - c\hat{\alpha}/(2\hat{\lambda})$. Indeed, it follows from the above-mentioned relation that this shift with arbitrary velocity $c$ gives rise to the expected shift of the velocity of the traveling pattern, $\hat{\mu} \to \hat{\mu} - c$.

*Long wave limit*: A straightforward consequence of eq. (11) is the solitary wave limit ($k \to 1$)

$$\Psi = \sqrt{A_0}\,\mathrm{sech}(x-\mu t)\exp\left\{\frac{i\mu x}{2\hat{\lambda}} - i\left[\frac{\mu^2}{4\hat{\lambda}} - \frac{\hat{\lambda}}{4}\right]t - \frac{i\hat{\alpha}A_0}{4\hat{\lambda}}\sin^{-1}[\tanh(x-\mu t)]\right\}, \tag{13}$$

subject to the first two constraints of eq. (12). It is seen that, for the modulus of the envelope to go like the square root of a hyperbolic secant, the wave must be appropriately chirped. Indeed, the envelope soliton in the form of the square root of a hyperbolic secant is typical for the quintic NLS equation.

(ii) The function

$$\Sigma = A_0^2 \mathrm{dn}^2(x - \mu t)$$

satisfies

$$\dot{\Sigma}^2 = 4(k^2 - 1)A_0^2 \Sigma + 4(2 - k^2)\Sigma^2 - \frac{4\Sigma^3}{A_0^2}.$$

Alignment with eq. (7) delivers another periodic solution, namely

$$\Psi = A_0 \mathrm{dn}(x - \mu t)\exp\left\{\frac{i\mu x}{2\hat{\lambda}} - i\left[\frac{\mu^2}{4\hat{\lambda}} - \hat{\lambda}(2 - k^2)\right]t - \frac{i\hat{\alpha} A_0^2}{4\hat{\lambda}}\int^{x-\mu t}\mathrm{dn}^2(\xi)\,d\xi\right\}, \qquad (14)$$

where the constraints are now given by

$$\hat{v} = -\frac{3\hat{\alpha}^2}{16\hat{\lambda}}, \quad \frac{\hat{\mu}}{\hat{\lambda}} - \frac{\hat{\alpha}\mu}{2\hat{\lambda}^2} = \frac{2}{A_0^2}. \qquad (15)$$

It is noted that the dispersion and quintic nonlinearity must have opposite signs.

*Long wave limit*: The solitary wave limit ($k \to 1$) yields

$$\Psi = A_0 \text{sech}(x - \mu t)\exp\left\{\frac{i\mu x}{2\hat{\lambda}} - i\left[\frac{\mu^2}{4\hat{\lambda}} - \hat{\lambda}\right]t - \frac{i\hat{\alpha} A_0^2}{4\hat{\lambda}}\tanh(x - \mu t)\right\}, \quad (16)$$

subject to the constraints in eq. (15). The wave must be appropriately chirped for the envelope to assume this form with hyperbolic secant amplitude.

It is noticed that setting $\hat{\alpha} = 0$ in eq. (15)$_1$ forces $\hat{v}$ to vanish. Thus, a purely hyperbolic secant profile is not compatible with the cubic-quintic nonlinear Schrödinger equation without self-steepening.

### Fronts

A front (or kink) constitutes a sharp transition from a constant plane wave to a distinct, but otherwise arbitrary, asymptotic state. A simple analytical model is given in the present context by

$$\Sigma = 1/[1 + C_0 \exp(-\beta(x - \mu t))],$$

which satisfies ($\beta$ = a real parameter)

$$\dot{\Sigma}^2 = \beta^2 \Sigma^2 (1 - \Sigma)^2,$$

where $\beta$ is a real parameter. Comparison with eq. (7) produces the front solution

$$\Psi = \left[\frac{1}{1+C_0 \exp(-\beta(x-\mu t))}\right]^{1/2} \exp\left\{\frac{i\mu x}{2\hat{\lambda}} - i\left[\frac{\mu^2}{4\hat{\lambda}} - \frac{\hat{\lambda}\beta^2}{4}\right]t \right.$$
$$\left. - \frac{i\hat{\alpha}\log[C_0 + \exp(\beta(x-\mu t))]}{4\beta\hat{\lambda}}\right\}, \tag{17}$$

where the constraints are now

$$-\frac{4\hat{v}}{3\hat{\lambda}} - \frac{\hat{\alpha}^2}{4\hat{\lambda}^2} = \beta^2, \qquad \frac{\hat{\alpha}\mu}{2\hat{\lambda}^2} = \frac{\hat{\mu}}{\hat{\lambda}} - \beta^2. \tag{18}$$

If self-steepening is absent ($\hat{\alpha} = 0$), eq. (17) degenerates to a kink studied earlier in the literature,[24] where the quintic nonlinearity and dispersion are of opposite signs.

### Dark Solitons

A dark soliton can, in principle, be obtained as a long wave limit of a sn periodic wave. Alternatively, starting directly with

$$\Sigma = A_0^2 \tanh^2(x - \mu t),$$

which obeys

$$\dot{\Sigma}^2 = 4A_0^2 \Sigma - 8\Sigma^2 + \frac{4\Sigma^3}{A_0^2},$$

one obtains the dark solitary pulse

$$\Psi = A_0 \tanh(x - \mu t) \exp\left\{ i\left(\frac{\mu}{2\hat{\lambda}} - \frac{\hat{\alpha} A_0^2}{4\hat{\lambda}}\right)x - i\left[\frac{\mu^2}{4\hat{\lambda}} + 2\hat{\lambda} - \frac{\hat{\alpha}\mu A_0^2}{4\hat{\lambda}}\right]t \right.$$
$$\left. + \frac{i\hat{\alpha} A_0^2 \tanh(x - \mu t)}{4\hat{\lambda}} \right\}, \qquad (19)$$

where

$$\hat{\nu} = -\frac{3\hat{\alpha}^2}{16\hat{\lambda}}, \qquad \frac{\hat{\alpha}\mu}{2\hat{\lambda}^2} - \frac{\hat{\mu}}{\hat{\lambda}} = \frac{2}{A_0^2}. \qquad (20)$$

Here, the quintic nonlinearity ($\hat{\nu}$) and dispersion ($\hat{\lambda}$) need to be of opposite sign and again a proper chirp must be present.

**The relation to the bright solitons in the absence of the self-steepening**

It is instructive to see how a particular known solution in the literature can be retrieved via the present algorithm. We restrict attention to the purely quintic case $\hat{\alpha} = 0$. Consider the case with

$$\Sigma = \frac{A_0^2}{1 + \beta \cosh[2\gamma(x - \mu t)]}, \qquad (A_0, \beta, \gamma \text{ real}) \qquad (21)$$

whence

$$\dot{\Sigma}^2 = 4\gamma^2 \left[ \Sigma^2 - \frac{2\Sigma^3}{A_0^2} + \frac{(1 - \beta^2)\Sigma^4}{A_0^4} \right].$$

Comparison with eq. (7) produces the solution for the cubic-quintic case without self-steepening,[16,17] namely,

$$\Psi = \frac{A_0 \operatorname{sech}[\gamma(x-\mu t)]\exp\left[\dfrac{i\mu x}{2\hat{\lambda}} + i\left(\gamma^2\hat{\lambda} - \dfrac{\mu^2}{4\hat{\lambda}}\right)t\right]}{\left\{2\beta + (1-\beta)\operatorname{sech}^2[\gamma(x-\mu t)]\right\}^{1/2}}. \qquad (22)$$

The amplitude factor $A_0$ and the parameter $\beta$ are given by

$$A_0^2 = \frac{4\gamma^2\hat{\lambda}}{\hat{\mu}}, \qquad \beta^2 = \frac{\hat{v}\,A_0^4}{3\hat{\lambda}\gamma^2} + 1, \qquad (23)$$

while the wave number $\gamma$ and the phase speed $\mu$ are related to the terms in the exponential phase factor of eq. (2) by

$$\gamma^2 = \frac{\mu v}{\hat{\lambda}} - \frac{\mu^2}{4\hat{\lambda}^2} - \frac{\lambda}{\hat{\lambda}}. \qquad (24)$$

The second constraint in eq. (23) shows why a hyperbolic secant profile is incompatible with the cubic-quintic NLSE. This would require $\beta = 1$ in eq. (21), whence $\hat{v} = 0$ and we recover the familiar hyperbolic secant profile of the cubic NLS equation.

## 4 Numerical Tests for the Stability of the Exact Solutions

The physical importance of the propagating patterns presented here requires that there be broad parameter ranges where such patterns are stable. We have performed systematic numerical simulations which demonstrate the existence of robust wave-propagation patterns. As concerns the computational techniques, a pseudospectral method in the spatial ($x$) direction is employed, while a fourth-order Runge-Kutta scheme with adaptive step size control is implemented in the time ($t$) domain. Two types of simulations have been conducted:

(A) Firstly, using perturbed patterns as initial conditions, sturdy propagation patterns were shown to persist in selected parameter regions, despite the presence of reasonably large initial random perturbations. On the other hand, some exact solutions were found to be unstable, suffering distortion or disintegrating upon the introduction of small perturbations.

More precisely, a perturbed initial condition was taken as

$\Psi|_{t=0} = (1 + \text{noise}) \Psi_{\text{exact}}|_{t=0}$, noise $= 0.1[1 - 2(\text{Rand})]$, where $\Psi_{\text{exact}}$ is an exact solution given by eqs. (13) or (16), and the 10% noise was generated by the standard random numerical variable 'Rand' in the interval of (0, 1).

(B) Secondly, eq. (1) was integrated forward in time directly from fairly arbitrary initial conditions. Robust solitary pulses may emerge, proving that these pulses are definitely stable solutions, as discussed in point (A) above, provided that the proper parameter regimes are chosen.

Here, we focus on the localized modes, and attention is restricted to certain interesting cases.

(A) *Initially perturbed exact wave profiles*

We first use the hyperbolic secant profile eq. (16) subject to the constraints given by eq. (15) to illustrate the dependence on amplitude. Secondly, we use the square root of hyperbolic secant profile eq. (13) subject to the constraints eq. (12) to show the effects of the quintic nonlinearity. In the terminology of temporal pulses in optical fibers, we select the 'anomalous' dispersion regime ($\hat{\lambda} > 0$) and positive cubic (Kerr) nonlinearity ($\hat{\mu} > 0$).

Starting with eq. (16) (a hyperbolic secant profile with an appropriate chirp), if one assumes $\hat{\lambda} > 0$, eq. (15) dictates the necessity of having a negative quintic nonlinearity ($\hat{v} < 0$). Thus, the constraint given by eq. (15) effectively defines a nonlinear dispersion relation between the speed $\mu$ and the amplitude parameter $A_0$. We first look at a special configuration where the wave is stationary ($\mu = 0$). Figure 1 shows that the disturbed solitary pulse quickly relaxes back toward the original shape and remains stable. For a propagating pattern, the velocity can be positive or negative, depending on the sign of the 'self-steepening' term $\hat{\alpha}$. Figure 2 shows that propagating pulse is robust, provided that the amplitude is sufficiently small. For larger amplitude, the pulse may become unstable and suffer disintegration (Figure 3).

Next we look at the square root of hyperbolic secant profile, eq. (13) with constrains eq. (12). Here, even with $\hat{\lambda} > 0$, $\hat{\mu} > 0$, the quintic nonlinearity can be of either sign (eq. (12)). For negative quintic nonlinearity ($\hat{\nu} < 0$), the pulse remains mostly stable (Figure 4). However, the stability property is less robust than that in the case for hyperbolic secant eqs. (16, 15), as a mild form of weak radiation/instability is observed, which potentially can destroy the pulse for large time. For positive quintic nonlinearity ($\hat{\nu} > 0$), a class of pulsating states / breathers is admissible (Figure 5). The amplitude varies periodically with time. However, there are also parameter regimes of strong instability, where the pulse disintegrates (Figure 6).

(B) *Arbitrary localized initial conditions*

We adopt the Gaussian as a typical example in this subsection, i.e., we replace the hyperbolic secant by a Gaussian in eq. (16) as the initial condition:

$$\text{sech}(z) \to \exp(-z^2/2), \tag{25}$$

and integrate eq. (1) forward in time by means of the same marching scheme as before. It is instructive to examine eq. (1) for the same parameter regions as in subsection (A) above, and to compare the properties in terms of the evolution for large time $t$.

For the parameter values corresponding to Figure 1, a localized mode still emerges, but the mode now propagates with a finite velocity (Figure 7).

Similarly, for the parameter values corresponding to those of Figure 2, where the solitary pulse is stable to a 10% noise, the numerical integration of an initially Gaussian profile eq. (25) still generates localized modes as the result of the evolution (Figure 8).

On the other hand, for 'marginally' stable modes, where weak instability/radiation is observed (Figure 4), the evolution of a Gaussian profile in those parameter regions leads to a splitting, distortion or strong perturbation of the localized pulse (Figure 9). Likewise, the evolution of a Gaussian profile in the parameter regime of the pulsating 'breather' also destroys its structure (Figure 10). Similar results are obtained for other initially localized profiles, e.g. an algebraically decaying one.

From these results, it is reasonable to conjecture that, in parameter regimes where the localized modes are stable, solitary pulses may emerge from the evolution of fairly arbitrary initial conditions.

## 5 Conclusions

A procedure for the calculation of propagating patterns for a class of DNLS equations with quintic nonlinearity[40–42] is presented. This method of constructing exact solutions exploits a pair of integrals of motion obtained via a wave packet ansatz.[43] Here, the evolution of the amplitude and phase of the wave envelope are reduced to standard canonical forms satisfied by elliptic functions. Explicit analytic expressions are obtained for, notably, periodic patterns, bright pulses, fronts or kinks and dark solitons. Illustrative numerical tests for stability of localized solutions are described.

The wave packet representation adopted here may also be applied to other DNLS equations. In fact, traveling waves for the DNLS equations incorporating a de-Broglie Bohm type quantum potential have been computed via an analogous procedure to that of the present paper.[43] Extensions to other physically interesting settings, e.g. scattering,[44] and hydrodynamic waves,[45–47] are amenable to this method. Thus, for NLS equations incorporating higher order derivative terms in $\Psi$ such as $\Psi_{xxxx}$, a pair of coupled nonlinear equations for $\phi$, $\psi$ is again obtained which admit a pair of invariants of motion. The application of the procedure in this case is under current investigation.

## Acknowledgements

Partial support has been provided by the Research Grants Council of Hong Kong. B.A.M. appreciates hospitality of the Department of Mechanical Engineering at the University of Hong Kong.

**Figures Captions**

(1) Figure 1 (Color online): Stability of a stationary solitary pulse described by eq. (16) (hyperbolic secant profile), $\hat{\lambda} = 1/2$, $\hat{\alpha} = 1$, $\hat{\mu} = 1$, $\hat{\nu} = -3/8$, $A_0 = 1$, $\mu = 0$ (pulse at rest). The perturbed profile quickly relaxes back to the exact, stationary solitary pulse.

(2) Figure 2 (Color online): Stability of a propagating solitary pulse described by eq. (16) (hyperbolic secant profile), $\hat{\lambda} = 1/2$, $\hat{\mu} = 1$, $\hat{\nu} = -3/8$, $A_0 = 1/2$, (a) $\hat{\alpha} = 1$, $\mu = -3$ (pulse moving to the left); (b) $\hat{\alpha} = -1$, $\mu = 3$ (pulse moving to the right). The 10% noise added initially does not affect the propagation of the pulses.

(3) Figure 3 (Color online): Instability of a pulse with a sufficiently large amplitude (eq. (16), hyperbolic secant profile), $\hat{\lambda} = 1/2$, $\hat{\alpha} = 1$, $\hat{\mu} = 1$, $\hat{\nu} = -3/8$, $A_0 = 2$, $\mu = 3/4$. Even with no noise added initially, the pulse disintegrates on numerical marching forward in time.

(4) Figure 4 (Color online): Evolution of a solitary pulse with cubic and quintic nonlinearities of opposite signs (eq. (13), the profile in the form of the square root of the hyperbolic secant), for $\hat{\lambda} = 1/2$, $\hat{\mu} = 1$, $\hat{\nu} = -7/20$, $A_0 = (15)^{1/2}$, and (a) $\hat{\alpha} = $ 1, $\mu = 1$ (pulse moving to the right), or (b) $\hat{\alpha} = -1$, $\mu = -1$ (pulse moving to the left).

(5) Figure 5 (Color online): Formation of a pulsating state from an initial condition described by eq. (13) (square root of hyperbolic secant profile), $\hat{\lambda} = 1/2$, $\hat{\mu} = 1$, $\hat{v} = 8$, $A_0 = (3/67)^{1/2}$, (a) $\hat{\alpha} = 1$, $\mu = 1$ (pulse moving to the right), (b) $\hat{\alpha} = -1$, $\mu = -1$ (pulse moving to the left). The amplitude of the pulse varies periodically with time.

(6) Figure 6 (Color online): Disintegration of a solitary pulse with cubic and quintic nonlinearities of the same sign (eq. (13), the profile in the form of the square root of the hyperbolic secant), for $\hat{\lambda} = 1/2$, $\hat{\mu} = 1$, $\hat{v} = +7/20$ (compare with Figure 4), $A_0 = (15/29)^{1/2}$, $\hat{\alpha} = 1$, $\mu = 1$.

(7) Figure 7 (Color online): A robust pulse with the same parameter values as in Figure 1, except that the profile is initially Gaussian (see eq. (25)).

(8) Figure 8 (Color online): Robust pulses for the same parameter values as in Figure 2, except that the profile is initially Gaussian (see eq. (25)).

(9) Figure 9 (Color online): Strong perturbations of marginally stable pulses at the same parameter values as in Figure 4, except that the profile is initially Gaussian (see eq. (25)).

(10) Figure 10 (Color online): Distortion / destruction of the breather at the same parameter values as in Figure 5, except that the profile is initially Gaussian (see eq. (25)).

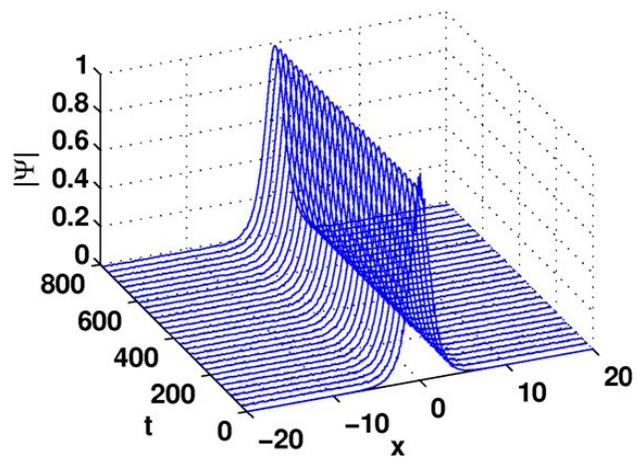

**C. Rogers et al    Figure 1**

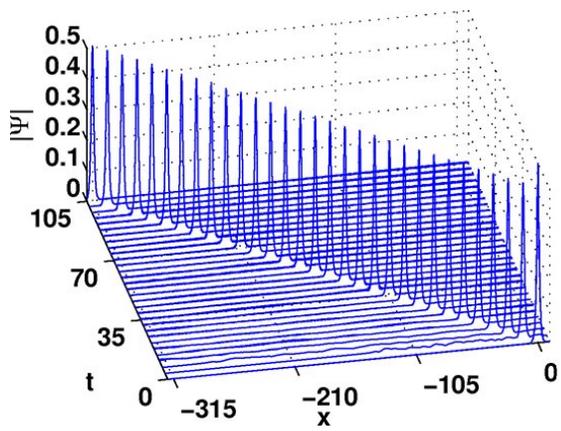 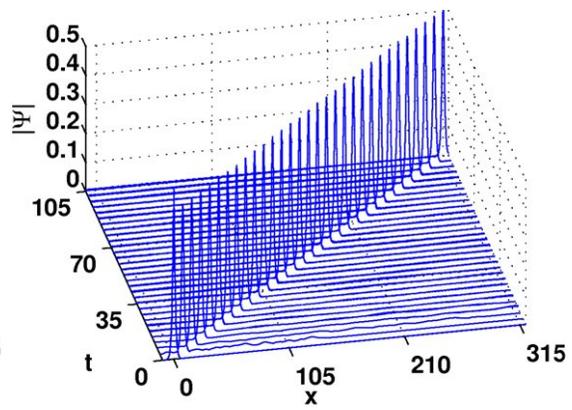

(a)          (b)

C. Rogers et al     Figure 2

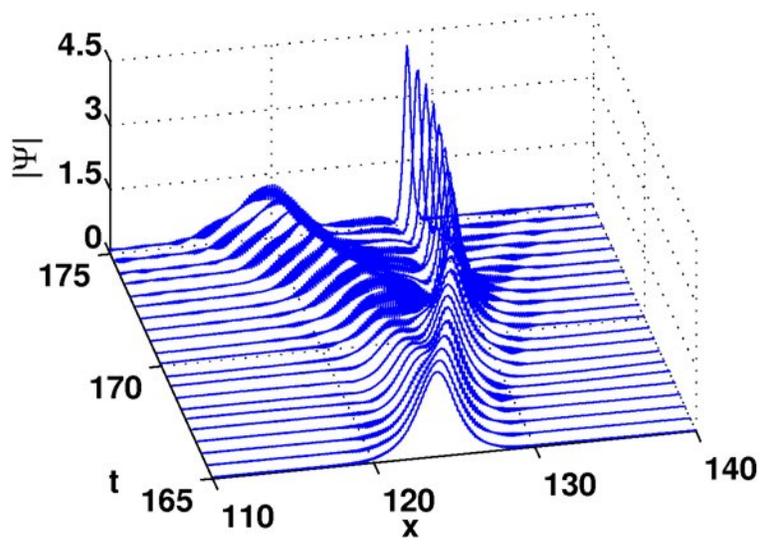

**C. Rogers et al    Figure 3**

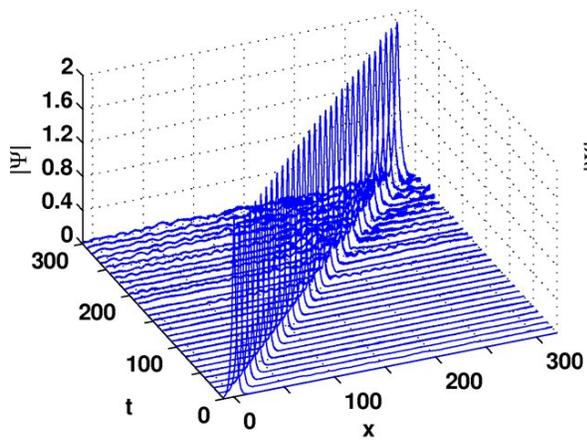 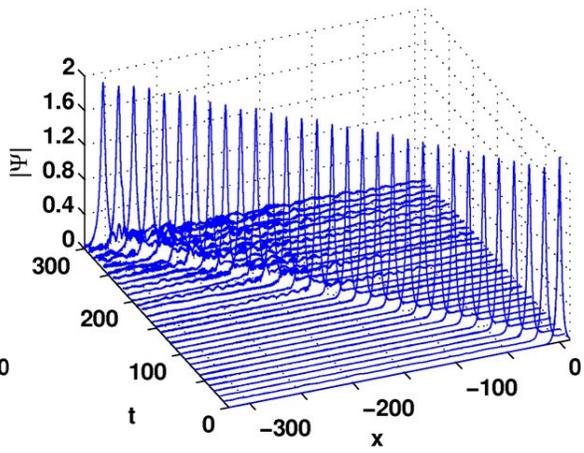

**(a)** **(b)**

**C. Rogers et al     Figure 4**

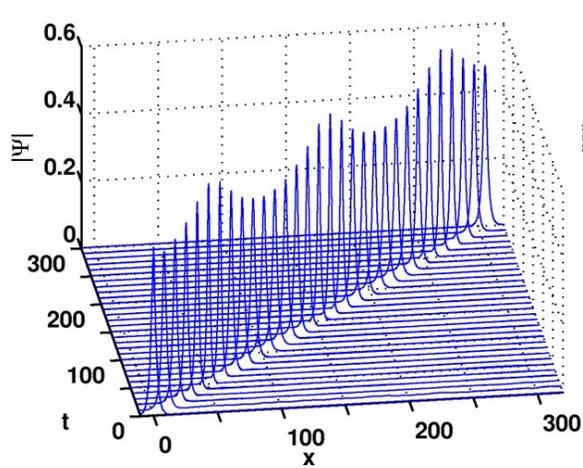 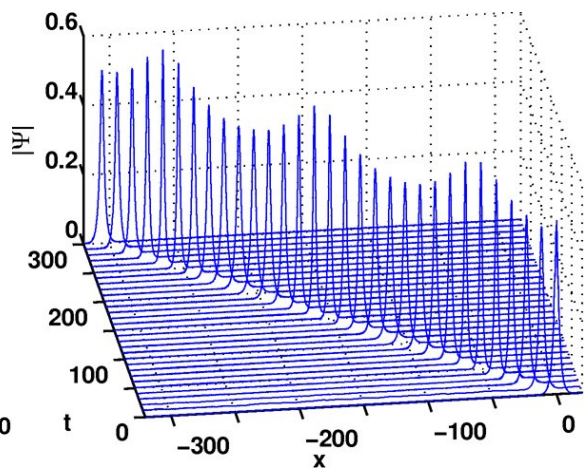

**(a)** **(b)**

**C. Rogers et al    Figure 5**

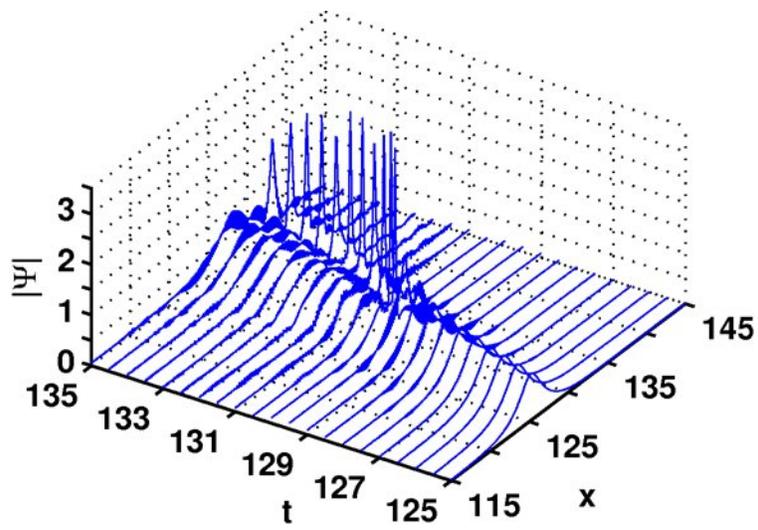

**C. Rogers et al    Figure 6**

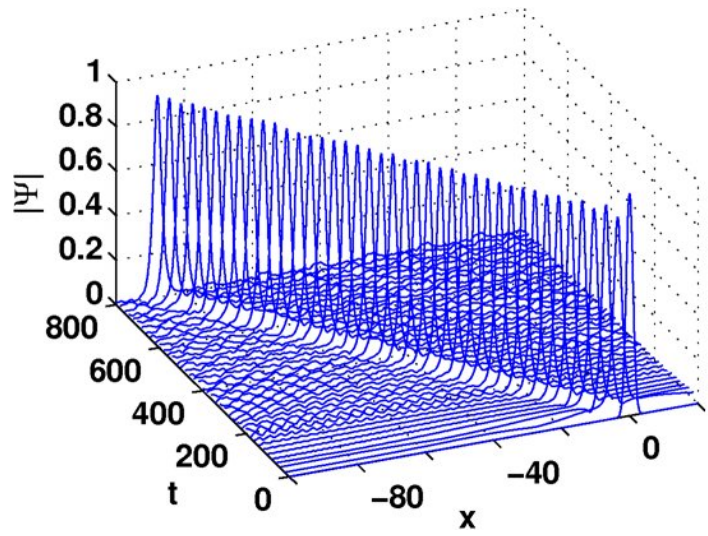

**C. Rogers et al**     **Figure 7**

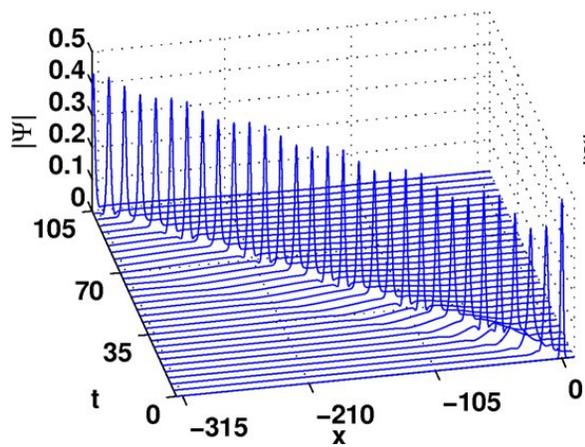
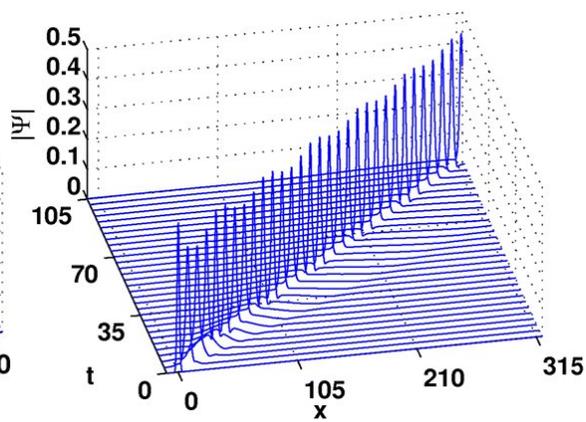

(a)            (b)

**C. Rogers et al     Figure 8**

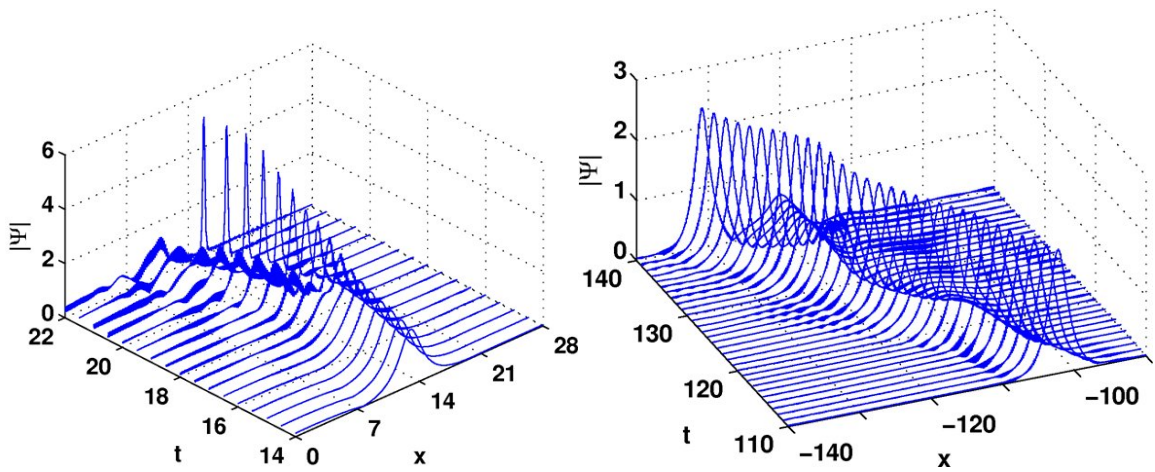

        **(a)**                                         **(b)**

**C. Rogers et al        Figure 9**

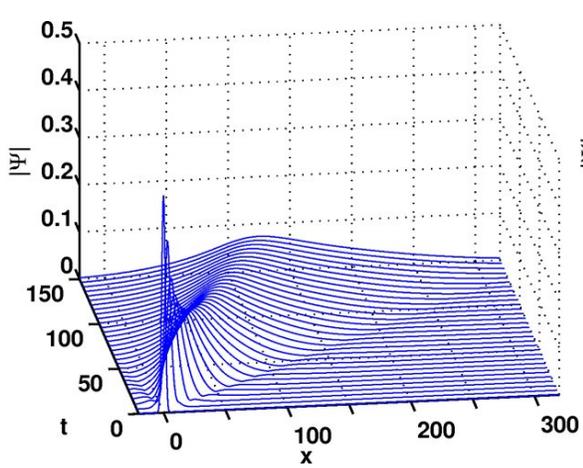 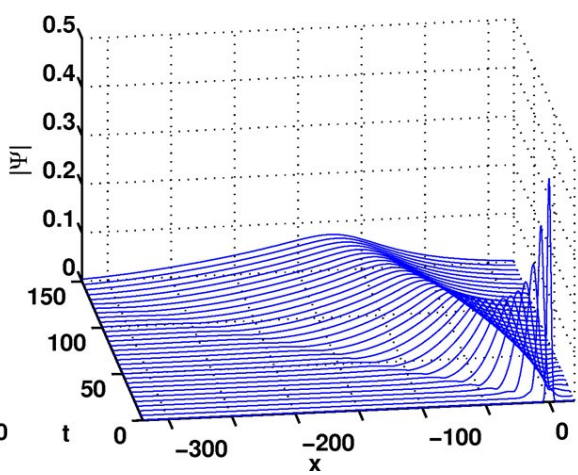

(a)          (b)

C. Rogers et al     Figure 10